\documentclass[12pt]{article}
\usepackage{graphics}
\begin{document}
\begin{flushright}
WISC-MILW-98-TH-18 \\
accepted for publication in {\em Physics Letters B}\\
\end{flushright}
\vspace{.5in}
\begin{center}
{\Huge Some Connections between \vspace{4mm}\\ Quantum Tunneling and Inflation}\\
\vspace{.25in}
{\Large John W. Norbury}\\
\vspace{.15in}
{\em Physics Department, University of Wisconsin - Milwaukee,}\\
{\em P.O. Box 413, Milwaukee, Wisconsin 53201, USA.}\\ (email: norbury@csd.uwm.edu)\\
\end{center}
\vspace{5mm} \noindent

\noindent
The Wheeler-DeWitt equation in the minisuperspace approximation is studied in four
different models. Under certain circumstances each model
leads to a tunneling potential and under the  same
circumstances the classical version of each model leads to inflation.

\vspace{10mm}
\vspace{10mm}

\newpage

One of the most important ideas in cosmology is the inflationary universe scenario
\cite{Olive90, Lidsey97} in which the universe underwent extremely rapid expansion with
positive acceleration at early times. Inflation solves some of the classic problems of
standard big bang cosmology having to do with the horizon, flatness, magnetic monoploes and
density fluctuations \cite{Kolb90}. Their solution in standard big bang cosmology
involved postulating special initial conditions, which can subsequently be explained within
the inflationary model. Recently cosmologists have become interested in the actual `birth'
of the universe, and one idea that has received attention is that the universe arose due to a
quantum tunneling process.  Many authors \cite{Hawking} have considered the connection between inflation and
quantum tunneling by calculating the tunneling wave function  and showing that it corresponds to an
inflating universe. In the present paper  the connection between inflation and
quantum tunneling is also considered, but from a different point of view. The wave function is not calculated,
but rather  the general properties of the tunneling potential are considered.  For the special cases
considered herein, it is found that the conditions for the potential to have a tunneling shape are the same
as the conditions for inflation.

\underline {\it 1. Cosmological Constant Model.}
In the Friedmann - Robertson - Walker (FRW) model, the
 expansion rate $\dot{a}$  of  the Universe
 is given by 
\begin{equation}
H^2 \equiv (\frac{\dot{a}}{a})^2=\frac{8\pi G}{3}\rho-\frac{k}{a^2}+\frac{\Lambda}{3}
\label{1-1}
\end{equation}
where $a$ is the scale factor, $H$ is the Hubble parameter, $\rho $ is the density of
matter or radiation in the Universe, $\Lambda $ is the cosmological constant and $k$ is the
curvature parameter. The acceleration $\ddot{a}$ is given by
\begin{equation}
 \frac{\ddot{a}}{a}=-\frac{4\pi G}{3}(\rho + 3p)+\frac{\Lambda}{3}
\label{1-2}
\end{equation}
 where  $p$ is the pressure  parameterized
as $p=\frac{\gamma}{3}\rho  $ where $\gamma = 0$ for  matter 
and $\gamma = 1$ for radiation. The conservation
law is
\begin{equation}
\frac{d}{da} (\rho a^{3+\gamma }) =0 .
\label{1-4a}
\end{equation}
If the cosmological constant $\Lambda $
dominates the right hand side of (\ref{1-1}) as 
$H^2 \equiv (\frac{\dot{a}}{a})^2=\frac{\Lambda}{3}$ then one obtains the exponential
inflationary solution
\begin{equation}
a(t) = a_0 e^{Ht}
\label{1-6}
\end{equation}
with $H\equiv \sqrt{ \frac {\Lambda}{3}}$.

Quantizing  equation (\ref{1-1}) gives the Wheeler-DeWitt equation in the
minisuperspace approximation \cite{Kolb90} as
\begin{equation}
[-\frac{d^2}{d a^2} + U(a) ] \Psi =0
\label{1-7}
\end{equation}
where the potential for a closed ($k=+1$) empty ($\rho =0$) universe  is 
\begin{equation}
U(a) = \frac{9\pi ^2}{4G^2} (a^2 - \frac{\Lambda }{3} a^4)
\label{1-8}
\end{equation}
This is plotted \cite{Kolb90} in Figure 1 which goes
like
$a^2$  for small $a$. This quadratic dependence changes at intermediate $a$ and the
potential reaches a maximum and then turns over due to the $ - \frac{\Lambda }{3} a^4$
term. The basic idea of quantum tunneling is that the universe began with a value $a =
0$ with zero energy and the potential barrier prevented expansion of the classical universe
by causing it to immediately re-collapse. However, quantum mechanically there is a
probability that the universe can tunnel through the  barrier and end up on the other
side with a non-zero value of $a$.  This value of $a$ would then keep getting bigger
\cite{Hawking}.

The above results are well known \cite{Kolb90} and serve partly as
introduction to what follows. However,  the point is that the  same  cosmological
constant $\Lambda $ that gives rise to inflation  also produces quantum tunneling when
the classical inflation equation is quantized. (It does not matter that the 
$\frac{k}{a^2} = \frac{+1}{a^2}$ term has been neglected in the classical equation
because when it is included it goes to zero anyway due to the huge size of $a^2$ in
inflation, producing an effectively flat universe. Also the cases for $k=0$ and $k=-1$ do not lead to
tunnelling. These cases are not discussed because they are unphysical in the sense that the correpsonding
action is infinite in the FRW model.) 
Models of open inflation \cite{Gott82, Ratra94}  are still consistent with the creation of a closed
universe,  because a closed bubble, formed in the process of false vacuum decay, looks like an open universe
from the inside of the closed bubble \cite{Linde95, Coleman80}.

\underline  {\it 2. Decaying Cosmological Constant Model.} It is not  easy to understand why the 
present day value of the cosmological
constant should be close to zero. Theory predicts a value  $10^{120}$ times bigger
than the experimental lower bound. This discrepancy is known as the ``cosmological constant problem"
\cite{Wei89}. Therefore many authors have considered theories in which the cosmological constant gets
smaller as $a$ gets bigger \cite{Ozer}.

If the right hand side of (\ref{1-1}) is dominated by a term of the form
$\frac{C}{a^m}$, as in $(\frac{\dot{a}}{a})^2=\frac{C }{a^m}$ then the solution (for $m \neq 0 $ ) is 
\begin{equation}
a(t) \propto t^{2/m}
\label{2-1}
\end{equation}
For instance if radiation density dominates the right hand side of (\ref{1-1}) then $C
\equiv \frac{8\pi G}{3} \rho_0 a_0^4$ and $m=4$.  If 
a decaying cosmological constant $\Lambda $ dominates
the right hand side of (\ref{1-1}) then
\begin{equation}
\Lambda \equiv \frac{3C}{a^m}
\label{2-2}
\end{equation}
Later it will be seen how scalar field theories can also be written in
the form of a decaying $\Lambda $.
Equation (\ref{2-1}) implies
\begin{equation}
\ddot a(t) \propto  (\frac{2}{m})(\frac{2}{m}-1)t^{\frac{2}{m}-2}
\label{2-3}
\end{equation}
for $m \neq 0 $.  Thus a matter ($m=3$)
or radiation ($m=4$) dominated universe corresponds to a universe which is 
decelerating, i.e. $\ddot a$ is negative. Usually one associates the idea of
inflation with the  exponential inflation behavior of (\ref{1-6}) which gives
 positive $\ddot a$. However inflation can occur via the
power law \cite{ Olive90,  Lidsey97, Lucchin85, Barrow87}  of equation (\ref{2-1}) provided
$\frac{2}{m}$ is sufficiently large. Actually a  general inflationary solution is
defined simply as one which gives the power of equation (\ref{2-1}) as $2/m > 1$. Thus
 general inflation occurs whenever
\begin{equation}
m < 2
\label{2-4}
\end{equation}
Substituting into (\ref{2-3}) means that general inflation is simply expansion with 
positive acceleration \cite{Lidsey97}.

If one simply inserts (\ref{2-2}) into (\ref{1-8}) the potential becomes
\begin{equation}
U(a) = \frac{9\pi ^2}{4G^2} (a^2 - C a^{4-m})
\label{2-5}
\end{equation}
By plotting this new $U(a)$ for various values of $m$, one finds that a  tunneling
potential (one that rises at small $a$, reaches a maximum and then falls off, similar to
Fig.1) will only occur for 
\begin{equation}
m < 2
\label{2-6}
\end{equation}
Thus the  quantum tunneling constraint (\ref{2-6}) is the same as  the requirement
(\ref{2-4}) for inflation. 

Actually the above simple argument is not strictly correct because the
conservation equation (\ref{1-4a}) is not valid for a changing cosmological constant. Instead the conservation
equation is modified to 
\begin{equation}
\frac{1}{a^{3+\gamma}} \frac{d}{da} (\rho a^{3+\gamma}) 
= - \frac{1}{8 \pi G} \frac{d\Lambda }{da}
\label{2-8}
\end{equation}
This shows that a decreasing (or increasing) cosmological constant results in production
(or absorption) of matter or radiation. Thus decaying cosmological constant models 
cannot be considered in vacuum ($\rho = 0$) as was done  above. One must explicitly include
the radiation or matter density ($\rho \neq 0$) which results from changing $\Lambda $. The Wheeler-DeWitt
equation must be re-derived. The potential $U(a)$ is then the same as (\ref{2-5}) except with an additional
density term. For example, if one initially specifies a radiation-type density of the form (with
$\Lambda \equiv 8 \pi G \rho_{v}$)

\begin{equation}
\rho + \rho_{v} \equiv \rho_{0}(\frac{a_0}{a})^4 + \rho_{v_0}(\frac{a_0}{a})^m
\label{2-8a}
\end{equation}
then the potential for a closed ($k=+1$) universe is

\begin{equation}
U(a)= \frac{9\pi^2}{4G^2} ( a^2- f a^{4-m} -b )
\label{2-9}
\end{equation}
where the constants are $f \equiv \frac{8\pi G}{3}\rho_{v_0} a_0^m  $ and $ b \equiv \frac{8\pi G}{3}\rho_{0}
a_0^4
$. Thus the simple  
argument of the previous paragraph remains intact.

 \underline  {\it 3. Scalar Field Model.}  The density and pressure for
an interacting scalar field $\phi $ is 
\cite{Kolb90}
\begin{equation}
\rho_\phi = \frac{1}{2} {\dot {\phi}}^2 + V(\phi)
\label{3-1}
\end{equation}
and 
\begin{equation}
p_\phi = \frac{1}{2} {\dot {\phi}}^2 - V(\phi)
\label{3-2}
\end{equation}
where $V(\phi )$ is the interaction potential. Assuming that the scalar field dominates,
 equation (\ref{1-1}) becomes
\begin{equation}
H^2 \equiv (\frac{\dot{a}}{a})^2=\frac{8\pi G}{3}[\frac{1}{2} {\dot {\phi}}^2 + V(\phi)]
\label{3-3}
\end{equation}
and instead of  (\ref{1-4a}) one obtains (assuming a massless field and ignoring
spatial derivatives) \cite{Kolb90}
\begin{equation}
\ddot \phi + 3 H \dot \phi + V^\prime = 0 
\label{3-4}
\end{equation}
which is the equation of motion for the scalar field where $V^\prime \equiv \frac{dV}{d\phi
}$. Equations (\ref{3-3}) and (\ref{3-4}) form a set of coupled equations. Substituting $H$ from (\ref{3-3})
into (\ref{3-4})  gives (with $k=\Lambda =0$)
\begin{equation}
\ddot{\phi}+\dot{\phi} \sqrt{12{\pi}G(\dot{\phi}^{2}+2V)}+V^\prime =0
\label{3-5}
\end{equation}
which can now be solved for $\phi (t)$ which can be put back into (\ref{3-3}) to obtain
$a(t)$. It is this $a(t)$ which will indicate whether inflation occurs. In particular,
 if the resulting $a(t)$ implies positive $\ddot a$ then  inflation does occur \cite{Lidsey97}. Substituting
$\rho$ and
$p$ from (\ref{3-1}) and (\ref{3-2}) into (\ref{1-2}) and ignoring $\Lambda $ gives
\begin{equation}
-qH^2 \equiv \frac{\ddot{a}}{a}=-\frac{8\pi G}{3}(\dot{\phi}^{2}-V)
\label{3-6}
\end{equation}
This shows that $\ddot a$ is positive (i.e. inflation occurs)  when \cite{Lidsey97}
\begin{equation}
V(\phi ) > \dot{\phi}^{2}  \hspace{10mm}   (\phi \neq constant)
\label{3-7}
\end{equation}
In the slow roll approximation $\dot \phi \approx 0$. Thus if $\phi = constant$ then this
condition (\ref{3-7}) for inflation reduces to the requirement only of a positive definite
potential, namely
\begin{equation}
V(\phi ) > 0  \hspace{10mm}   (\phi = constant)
\label{3-7a}
\end{equation}

The usual way of quantizing the FRW model for a scalar field is to treat $\phi $ and $a$ as
independent variables each with their own canonical momenta. The Wheeler-DeWitt equation is
then \cite{Kolb90}
\begin{equation}
[ -\frac{{\partial }^2}{\partial a^2} + \frac{3}{4 \pi G} \frac{1}{a^2} 
\frac{{\partial }^2}{\partial {\phi }^2} + U(a, \phi)] \Psi = 0
\label{3-8}
\end{equation}
where the potential for a closed ($k = +1$) universe is
\begin{equation}
U(a, \phi) = \frac{9 {\pi }^2}{4 G^2} [ a^2 - \frac{8 \pi G}{3} V(\phi ) a^4 ]
\label{3-9}
\end{equation}

Because $U$ depends on both $\phi$ and $a$ then tunneling can occur in either the $\phi$ or $a$ direction.
In this work only tunneling in the $a$ direction is considered, because it is this type
of tunneling that has been associated with the `birth' of the Universe \cite{Hawking}.

If $\phi = constant$ then $V(\phi )$ is also constant, just like the constant $\Lambda $ of Section {\it 1}.
If
$V(\phi ) $ is zero or negative at any fixed value of $\phi$ then $U(a, \phi)$ will be an infinitely rising
curve as a function of $a$ and will  not yield a tunneling potential (like Fig.1). For positive
definite $V(\phi ) > 0$  the potential
$U(a, \phi)$ in (\ref{3-9}) looks exactly like Fig.1 and thus tunneling  will occur
for  arbitrary $V(\phi ) > 0$. It was also found that inflation occurs for
arbitrary $V(\phi ) > 0$ in (\ref{3-7a}).  

This section has dealt {\em only} with the case where $\phi$ is constant.
 The case of variable $\phi$ is now
considered.

\underline  {\it 4. Uniform Scalar Density Model.} Now examine the case for $\phi \neq constant $.
The full Wheeler-DeWitt equation will not be solved, but rather a `semiclassical' model will be considered
in which the variation of $\phi$ is obtained using a classical argument. Also only one specific model for 
$V(\phi )$ will be considered.

In examining tunneling we are interested in the $a$ dependence of $U(a, \phi)$ given in (\ref{3-9}). Thus in
considering a variable $\phi$, we want its variation as a function of $a$.  Classically the evolution of the
scalar field
$\phi $ is tied to the evolution of $a$. This can be seen as follows. Upon specifying
$V(\phi ) $, equation (\ref{3-5}) is solved for
$\phi (t)$ which when substituted into (\ref{3-3}) enables a solution for $a(t)$ to be
found. Given $\phi (t)$ and $a(t)$ allows  $t$ to be eliminated giving
$\phi =
\phi (a)$. Using (\ref{3-1}) gives a function of $a$ namely $\rho_\phi = \rho_\phi (a)$. That is,
 the scalar field can be expressed in terms of a uniform density $\rho_\phi (a)$. In most
cases these steps will have to be done numerically because (\ref{3-5}) cannot be solved for
arbitrary
$V(\phi ) $. However now consider a model by Barrow  \cite{ Lucchin85, Barrow87} which can be
solved analytically. His potential is 
\begin{equation}
V(\phi)\equiv{\beta}e^{-\lambda\phi}
\label{4-1}
\end{equation}
where $\beta$ and $\lambda$ are constants to be determined.  
This potential is like the one studied by Ratra \cite{Ratra89} who gave a
perturbative solution to the Wheeler-DeWitt equation to calculate the spectrum of density 
fluctuations during
inflation. Barrow 
\cite{Barrow87} shows that a  particular solution to (\ref{3-5})
is 
\begin{equation}
\phi(t)=\sqrt{2A} \hspace{2mm} {\ln}t
\label{4-2}
\end{equation}
where $\sqrt{2A}$ is just some constant.
This claim is checked by substituting (\ref{4-1}) and (\ref{4-2}) into 
(\ref{3-5}).  From this it is found  that
\begin{equation}
\lambda=\sqrt{\frac{2}{A}}
\label{4-3}
\end{equation}
and 
\begin{equation}
\beta=-A
\label{4-4}
\end{equation}
or
\begin{equation}
\beta=A(24{\pi}GA-1)
\label{4-5}
\end{equation}
Barrow has a typing error when he writes $\lambda{A}=\sqrt{2}$. 
Also he uses units with $8\pi{G}=1$, so that the second solution (\ref{4-5}),
he writes  as $\beta=A(3A-1)$.  Barrow doesn't use the first
solution (\ref{4-4}) for reasons which will be seen shortly.

Having solved for $\phi(t)$, it is  now substituted into (\ref{3-3})
to solve for $a(t)$.   Substituting (\ref{4-2}) into (\ref{4-1}) gives $V=\frac{\beta }{t^{2}}$.
Substituting
this and $\dot{\phi}=\frac{\sqrt{2A}}{t}$ into (\ref{3-3}) one obtains
\begin{equation}
H^{2}\equiv(\frac{\dot{a}}{a})^{2}=\frac{8{\pi}G}{3}
(\frac{1}{2}\frac{2A}{t^{2}}+\frac{\beta}{t^{2}})=
\frac{8{\pi}G}{3}(A+\beta)\frac{1}{t^{2}}
\label{4-6}
\end{equation}
giving an equivalent density
\begin{equation}
\rho=\frac{A+\beta}{t^{2}}
\label{4-7}
\end{equation}
Clearly it is seen why  the first solution (\ref{4-4}) with 
$\beta=-A$ is rejected.  It would  give zero density.  Using the second solution 
(\ref{4-5}) with $\beta=A(24{\pi}GA-1)$ yields
\begin{equation}
\rho=\frac{24{\pi}GA^{2}}{t^{2}}.
\label{4-8}
\end{equation}
Solving  equation (\ref{4-6}) gives 
\begin{equation}
a =   F  t^{ 8{\pi}GA}
\label{4-9}
\end{equation}
where $F$ is a constant. Power law inflation results for 

\begin{equation}
8\pi G A > 1.
\label{4-11}
\end{equation}
Using $V=\frac{\beta}{t^2} = \frac{A(24\pi G A -1)}{t^2}$ and ${\dot{\phi}}^2 =
\frac{2A}{t^2}$, the condition
$8\pi G A > 1$ is seen to be equivalent to $V>{\dot{\phi}}^2$ in (\ref{3-7}).

Inverting the solution (\ref{4-9}) yields
\begin{equation}
t^{2}=(\frac{a}{F})^{\frac{1}{4\pi G A }}
\label{4-12}
\end{equation}
and substituting into (\ref{4-8}) gives
\begin{equation}
\rho = \frac{D}{a^{\frac{1}{4\pi G A }}}  \equiv  \frac{D}{a^m}
\label{4-13}
\end{equation}
where $D \equiv 24\pi G A^2 F^{\frac{1}{4\pi G A }}$. Thus  the scalar field density is equivalent to a
decaying cosmological constant. For the inflationary result, the condition, $8\pi G A >1$, gives
\begin{equation}
\frac{1}{4\pi G A } \equiv m < 2
\label{4-14}
\end{equation}
Thus it is  seen in equation (\ref{4-11}) that the model leads to
inflation for $8\pi G A>1$ which is seen to be consistent with the original $m<2$ constraint
in (\ref{2-4}) and the constraint $V>\dot \phi ^2$ in (\ref{3-7}).

When $\phi = constant$, it was shown that (\ref{3-9}) will always be a tunneling potential for arbitrary
positive definite $V(\phi)$, which was identical to the inflationary condition (\ref{3-7a}). However for 
$\phi \neq constant$ then $V(\phi )$ is not constant and could acquire a $t$ or $a$ dependence. This
can be examined by using the 
$
\phi (t)$ and
$a(t)$ solutions.   
 $a(t)$ is inverted to obtain $t(a)$ as in (\ref{4-12}). This is  substituted
 into $ \phi  (t)$ to obtain  $ \phi (a)$ which is put back into $V(\phi )$ to get
$V(a)$. 
Thus
\begin{equation}
V[\phi (a)] = \frac{K}{ a^\frac{1}{4\pi G A } }
\label{4-15}
\end{equation}
where $K\equiv \beta F^{\frac{1}{4\pi G A }}   $which gives (\ref{3-9}) as
\begin{equation}
U[a, \phi (a)] = \frac{9\pi^2}{4G^2} [a^2 - \frac{8\pi G}{3}  K a^{(4-\frac{1}{4\pi G A})} ]
\label{4-16}
\end{equation}
Tunneling will not be destroyed if
\begin{equation}
4-\frac{1}{4\pi G A} > 2
\label{4-17}
\end{equation}
which is  identical to the inflationary condition (\ref{4-11}). 

In summary, four different models have been studied and it has been found that the conditions for inflation
and quantum tunneling are the same for these four models. Future work will be devoted to considering non-FRW
models and to expanding the study of model 4 by seeing if the results hold more generally, without the use of
the `semiclassical' argument and with potentials more general than the Barrow model.

I am very grateful to Professor Leonard Parker for many useful conversations and for his comments on the
manuscript. This work was supported by the Wisconsin Space Grant Consortium.

\includegraphics{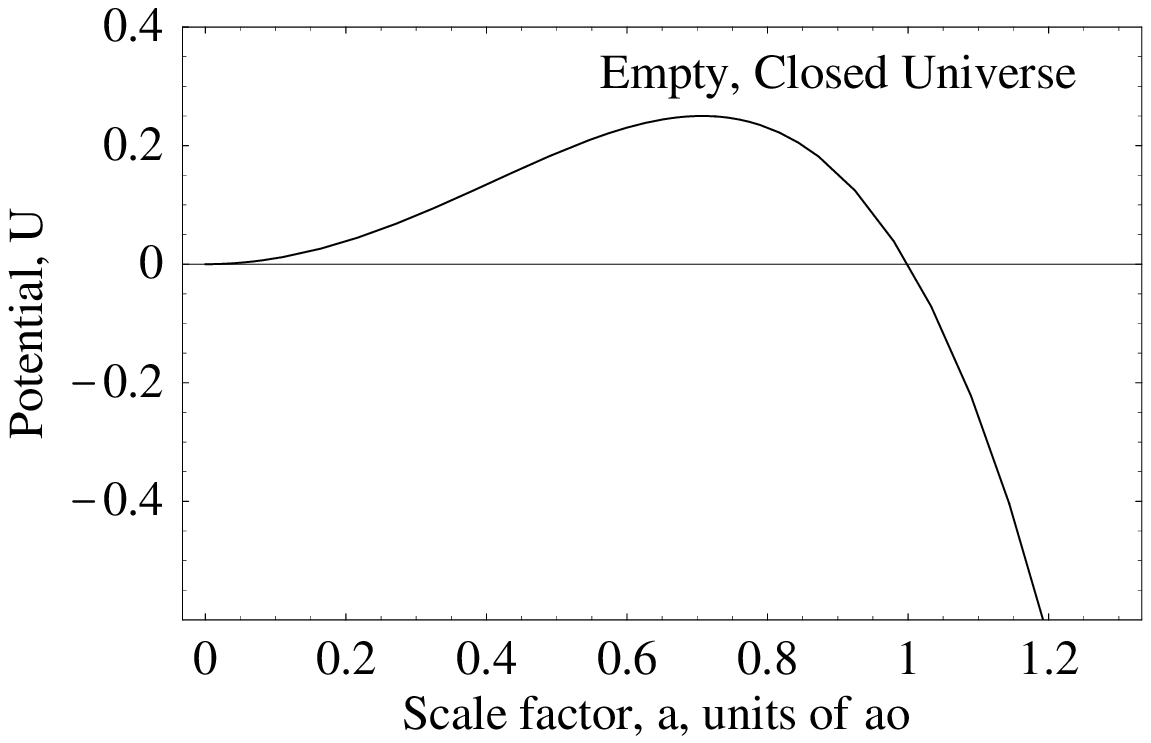}

\noindent
\hspace{50mm} {\bf Figure  1}\vspace{5mm}\\
Wheeler-DeWitt potential \cite{Kolb90} with cosmological constant. 
No matter or radiation is present. ($ao \equiv \sqrt{3/\Lambda}$)


\begin{thebibliography}{99}

\bibitem{Olive90} K. Olive, Phys. Rep. {\bf 190}, 307 (1990).

\bibitem{Lidsey97} J. Lidsey, et al, Rev. Mod. Phys., {\bf 69}, 373 (1997).

\bibitem{Kolb90} E. Kolb and M. Turner, {\em The Early Universe}, (Addison-Wesley, 1990).

\bibitem{Hawking}  S. Hawking,  Nucl. Phys. B {\bf 239}, 257 (1984); A. Vilenkin, Phys.
Rev. D {\bf 27}, 2848 (1983); W. Fischler, B. Ratra and L. Susskind, Nucl. Phys. B {\bf 259}, 730 (1985); I.
Moss and W. Wright, Phys. Rev. D {\bf 29}, 1067 (1984); G. Horowitz, Phys. Rev. D {\bf 31}, 1169 (1985); G.
Gibbons and L. Grishchuk, Nucl. Phys. B {\bf 313}, 736 (1989).

\bibitem{Gott82} J.R. Gott, Nature, {\bf 295}, 304 (1982).

\bibitem{Ratra94} B. Ratra and P.J.E. Peebles, Astrophysical Journal, {\bf 432}, L5 (1994).

\bibitem{Linde95} A. Linde, Phys. Lett. B, {\bf 351}, 99 (1995).

\bibitem{Coleman80} S. Coleman and F. De Luccia, Phys. Rev. D {\bf 21}, 3305 (1980).

\bibitem{Wei89} S. Weinberg, Rev. Mod. Phys.  {\bf 61}, 1 (1989).

\bibitem{Ozer}  
L.Parker and D.Toms,  Phys.Rev.D.{\bf 31}, 2424 (1985); 
L.Ford, Phys.Rev.D {\bf 35}, 2339 (1987); 
A.Dolgov, JETP Lett.{\bf 41}, 345 (1985);
M.Ozer and M.Taha, Nucl.Phys.B {\bf 287}, 776 (1987); 
J.Matyjasek,  Phys.Rev.D {\bf 51}, 4154 (1995);
P.Peebles and B.Ratra, Astrophys.J.{\bf 325},  L17 (1988); 
J.Lima, and J.Maia, Phys.Rev.D {\bf 49}, 5597 (1994); 
S.Barr, Phys.Rev.D {\bf 36}, 1691 (1987); 
M.Gasperini, Phys.Lett.B {\bf 194}, 347 (1987);
K.Freese, et al, Nucl.Phys.B {\bf 287}, 797 (1987);
M.Reuter and C.Wetterich, Phys.Lett.B {\bf 188}, 38 (1987); 
Y.Fujii and T.Nishioka, Phys.Rev.D {\bf 42}, 361 (1990); 
D.Pavon, Phys.Rev.D {\bf 43}, 375 (1991); 
M.Berman, Phys.Rev.D {\bf 43}, 1075 (1991);
A.Abdel-Rahman, Phys.Rev.D {\bf 45}, 3497 (1992); 
T.Olson and T.Jordan, Phys.Rev.D {\bf 35},  3258 (1987); 
W.Chen and Y.Wu, Phys.Rev.D {\bf 41}, 695 (1990);
K.Coble et al, Phys.Rev.D {\bf 55}, 1851 (1997).

\bibitem{Lucchin85} F. Lucchin and S. Matarrese, Phys. Rev. D {\bf 32}, 1316 (1985)

\bibitem{Barrow87} J.  Barrow, Phys. Lett. B, {\bf 187}, 12 (1987).

\bibitem{Ratra89} B. Ratra, Phys. Rev. D, {\bf 40}, 3939 (1989).
\end{thebibliography}
\end{document}